\documentclass{article}

\usepackage{arxiv}

\usepackage[utf8]{inputenc} 
\usepackage[T1]{fontenc}    
\usepackage{hyperref}       
\usepackage{url}            
\usepackage{booktabs}       
\usepackage{amsfonts}       
\usepackage{nicefrac}       
\usepackage{microtype}      
\usepackage{lipsum}		
\usepackage{natbib}
\usepackage{doi}
\usepackage{eurosym} 
\usepackage{graphicx} 
\usepackage{amssymb} 
\usepackage{amsmath}
\usepackage{url} 

\title{Towards a data-driven debt collection strategy based on an advanced machine learning framework}

\author{ \href{https://orcid.org/0000-0002-6542-2875}{\includegraphics[scale=0.06]{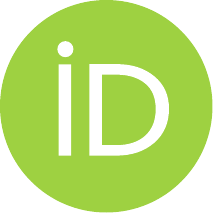}\hspace{1mm}Abel Sancarlos} \\
	\And
	\href{https://orcid.org/0000-0001-7837-1897}{\includegraphics[scale=0.06]{orcid.pdf}\hspace{1mm}Edgar Bahilo} \\
    \And 
    Pablo Mozo 
    \And
    Lukas Norman
    \And
    Obaid Ur Rehma
    \And 
    Mihails Anufrijevs
}

\renewcommand{\shorttitle}{Towards a data-driven debt collection strategy based on an advanced ML framework}

\hypersetup{
pdftitle={A template for the arxiv style},
pdfsubject={q-bio.NC, q-bio.QM},
pdfauthor={David S.~Hippocampus, Elias D.~Striatum},
pdfkeywords={First keyword, Second keyword, More},
}

\begin{document}
\maketitle

\begin{abstract}
The European debt purchase market as measured by the total book value of purchased debt 
approached  \euro 25bn in 2020 and it was growing at double-digit rates. 
This is an example of how big the debt collection and debt purchase industry has grown 
and the important impact it has in the financial sector. However, in order to ensure an 
adequate return during the debt collection process, a good estimation of the propensity 
to pay and/or the expected cashflow is crucial. These estimations can be employed, for instance, 
to create different strategies during the amicable collection to maximize quality standards 
and revenues. And not only that, but also to prioritize the cases in which a legal process 
is necessary when debtors are unreachable for an amicable negotiation. This work 
offers a solution for these estimations. Specifically, a new machine learning modelling 
pipeline is presented showing how outperforms current strategies employed in the sector. 
The solution contains a pre-processing pipeline and a model selector based on the best model calibration. Performance is validated with real historical data of the debt industry.
\end{abstract}


\section{INTRODUCTION} \label{sec:intro}

Nowadays, probability prediction models play an essential role in our modern lives and are in high demand in many industries. To cite only a few examples, these applications can range from models in the engineering sector, meteorology (weather forecasting), or online recommendation systems to modeling frameworks in the financial sector such as the prediction of the number of claims in insurance \citep*{Fissler2022caliasses,Kumar2019,dimitriadis2023triptych}, credit scoring or propensity to pay forecasts \citep*{rosch2020deep}. For these reasons, we should not be surprised to see that methods for evaluating and comparing probability forecasts are in increasing demand.  

Probability calibration is typically assessed graphically via reliability diagrams \citep*{Murphy1977, Broecker2007} that plot an estimated version of the conditional event probability (CEP) against the forecast value, with deviations from the diagonal suggesting lack of calibration. Classical approaches to estimating CEP rely on binning and counting and have been hampered by ad hoc implementation decisions, instability under unavoidable choices regarding binning, and inefficiency \citep*{Dimitriadis2021,ArrietaIbarra2022_metapaper,Roelofs2022} 

In addition, there has been a surge of interest in calibration metrics in the machine learning 
literature \citep*{dimitriadis2023triptych}. The popular and widely employed metric of the expected or estimated calibration 
error (ECE) \citep*{PakdamanNaeini2015,pmlr-v70-guo17a} depend on binning and counting and thus has important drawbacks to circumvent 
\citep*{Dimitriadis2021,Roelofs2022} and biases \citep*{Brocker2012,Ferro2012}. In fact, \cite{ArrietaIbarra2022_metapaper} follows up and elaborates on problems highlighted earlier by \citep*{gupta2021calibration,Roelofs2022} where they point out that the classical empirical calibration errors based on binning vary significantly based on the choice of bins. The choice of bins is fairly arbitrary and enables the analyst to fudge results (whether purposely or unintentionally).

Furthermore, \cite{ArrietaIbarra2022_metapaper} makes an extensive comparison between ECEs and ECCEs (empirical or estimated cumulative calibration errors) with proofs and mathematical results that emphasize the suitability of the latter (specifically, the ECCE-MAD and the ECCE-R). It is shown that the empirical cumulative calibration errors are fully non-parametric and uniquely, fully specified, statistically powerful and reliable. And not only that, but also without any specific trade-off concerning binning.

This paper applies the results of this last work of the ECCE-MAD and ECCE-R assessment to the debt collection industry, where having a good calibration of probabilities can be critical and can significantly lead to increased cash flows (e.g. in the hundreds of thousands or millions of euros). An advanced machine learning framework is designed to address the proper scoring rules for the Hyperparameter Optimization (HPO) and the use of the ECCE-MAD and ECCE-R to have the best model selection in terms of calibration.

This is crucial for this industry application. To have some orders of magnitude, as introduced in the abstract, the European debt purchase market as measured by the total book value of purchased debt approached \euro25bn in 2020 \citep*{debtreport}. This is an example of the important impact it has in the financial sector. The good estimation of the propensity to pay (PtP) is critical in order to ensure an adequate return during the debt collection process. Not only because of this quantity of interest (QoI) itself but also because of the derived QoIs from the PtP that can provide business value and drive decisions such as the expected payment amounts. These estimations can be employed, for instance, to create different strategies during the amicable collection to maximize quality standards and revenues. And not only that, but also to prioritize the cases in which a legal process is necessary when debtors are unreachable for an amicable negotiation.

The remainder of the article is organized as follows. Section \ref{sec:metrics} introduces the ECCE-MAD and the ECCE-R as well as the advantages of them according to previous research. Section \ref{sec:ml_proposal} and \ref{sec:app_des} presents the proposed advanced machine learning framework to ensure improved probability forecasts in the debt collection industry. Section \ref{sec:results} shows results for the specific debt collection application: a propensity-to-pay (PtP) model to drive strategies in collection strategy. Finally Section \ref{sec:conclusions} contains the conclusions of the work.

\section{METRICS OF CALIBRATION} \label{sec:metrics}

This section introduces some mathematical terminology as well as the advantages of cumulative metrics over binned metrics. The detailed results and proofs can be found in \cite{ArrietaIbarra2022_metapaper}.

Basically, the estimated/empirical calibration error ECE is the weighted average over each bin of the difference between the empirical probability and the mean predicted probability of each bin. This weighted average can be computed considering the number of bins or the bin lengths. In addition, the ECE can be defined in function of the $l^1$ or $l^2$ norm ($\text{ECE}^1$ and $\text{ECE}^2$ respectively). For the unfamiliar reader, we recommend reviewing the detailed definition in \cite{ArrietaIbarra2022_metapaper}.

Now, consider the following general setting. Suppose we have $n$ observations $y_1, y_2, \dots, y_n$ of the outcomes of independent Bernoulli trials with corresponding predicted probabilities of success, say $\tilde{y}_1, \dots, \tilde{y}_n$. For our purposes, each $\tilde{y}_k$ is the classifier's probabilistic score and the $y_k$ can be regarded as the binary class label. In addition, let's consider that the samples are sorted (preserving the pairing of $y_k$ with $\tilde{y}_k$ for every $k$) such that $\tilde{y}_1 < \tilde{y}_2 < \cdots < \tilde{y}_n$.
The cumulative differences are:
\begin{equation} \label{eq:cum_diff}
    C_k = \frac{1}{n} \sum_{j=1}^{k} (y_j - \tilde{y}_j)
\end{equation}
for $k = 1, 2,\dots, n$
Then, the maximum absolute deviation of the empirical cumulative calibration error (ECCE-MAD) is:
\begin{equation}
    \text{ECCE-MAD} = \max_{1 \leq k \leq n} \mid C_k \mid 
\end{equation}
and the range of the empirical cumulative calibration error (ECCE-R) is:
\begin{equation}
    \text{ECCE-R} = \max_{1 \leq k \leq n} C_k - \min_{1 \leq k \leq n} C_k
\end{equation}
where $C_k$ is defined in \eqref{eq:cum_diff} and $C_0=0$.

The following advantages motivates the choice of the ECCEs over the ECEs:
First, a trade-off\footnote{Widening the bins averages away more noise in the estimates, while
sacrificing some of the power to resolve variations as a function of score. Narrowing the bins resolves
finer variations as a function of score, while not averaging away as much noise in the estimates.} between 
statistical confidence and power to resolve variations as a function of score is inherent to the ECEs based on binning, while the ECCEs have no 
such explicit trade-off. In addition, obtaining meaningful ECEs requires
the sample size $n$ to be much larger
Secondly, the ECEs also exhibit an extreme dependence on the choice of bins, with different choices of bins 
yielding significantly different values for the ECE metrics; choosing among the possible binnings can be 
confusing, yet makes all the difference. In contrast, the ECCEs yield trustworthy results without needing such 
large numbers of observations and without needing to set any parameters.

\section{PROPOSED MACHINE LEARNING PIPELINE FOR AN IMPROVED PTP MODEL} \label{sec:ml_proposal}

\subsection{Introduction and Solution Description} \label{sec:desc_pl}

As presented in Section \ref{sec:intro}, there is a huge and increasing importance of the calibrated responses in the debt purchase and collection industry.

However, it is well-known that probability estimates from classifiers are often poorly calibrated \citep*{Loefstroem2023,euroscipy2023} thus requiring in those cases a calibration. In addition, as discussed in previous sections, the widely used metrics to assess calibration are far from being perfect affecting the reliability of an automatic model selection based on this assessment.

Having the best possible calibration in this business is essential, as it allows to assess more accurately and trustworthy the revenues of the different actions of the company. For example, in prioritizing the best subset of potential debtors to take legal actions based on the PtP or when assessing the number of payers in a debt portfolio prior to its purchase. And not only that, but a better calibration makes the models and the explanations better by accurately representing reality \citep*{Loefstroem2023} improving the explainability of the decisions taken from a business perspective.

For these reasons, the main priority for model selection in these settings is the good calibration of the probabilities. However, an automatic solution is required. This way, the training and monitoring processes of the solution  can be automated. In fact, this is a critical feature in the industry if you want to have a solution with minimal degradation (by constant updating), without manual intervention and without large human costs for constant evaluation. For this, a solution is outlined without relying on the visual judgment of graphs and with innovative proposed metrics (Section \ref{sec:metrics}) that behave better mathematically to guarantee the clear primary goal. See Section \ref{sec:metrics} to check the main advantages.

\begin{figure}[h] 
\centering
\vspace{.3in}
\includegraphics[scale=0.8]{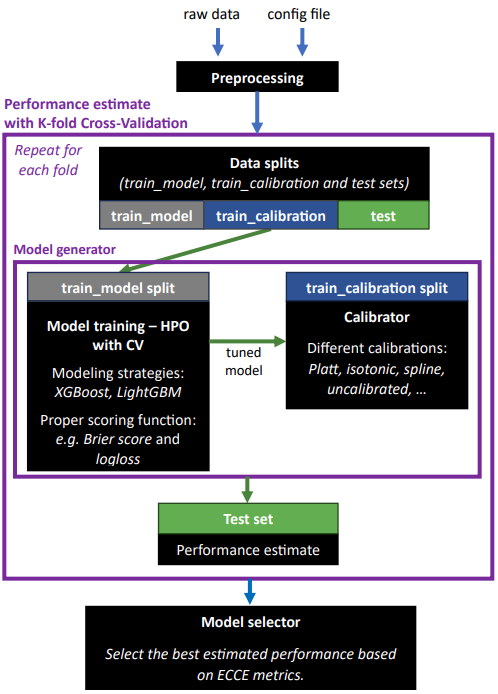}
\vspace{.3in}
\caption{Diagram of the proposed Machine Learning Pipeline for improved results in PtP and probability estimates.}
\label{fig:diagram}
\end{figure}

Figure \ref{fig:diagram} shows a diagram of the proposed advanced machine learning pipeline. The goal of this pipeline is to obtain the best possible PtP model to guide Operations Departments in the amicable collection strategy of the debt collection. Furthermore, this solution can also be valid for any probability model to be extracted in this sector.

Conceptually, the big picture of the strategy is as follows. First, we have a model generator that creates and tunes different models to be tested. Then, the performance of these models during the inference phase is estimated with K-fold Cross-validation. Finally, a model selector is applied based on the ECCE-MAD and ECCE-R to have the best model selection in terms of calibration, which is the primary goal for this industry application.

Looking in more detail at each of the parts (Figure \ref{fig:diagram}), we find the following steps:

\begin{enumerate}
    \item Raw data is preprocessed according to the config file.
    \item For each iteration of the outer CV loop, preprocessed data is split in three sets: train\_model, train\_calibration and test sets.
    \item To generate a model proposal, the model generator performs two steps:
    \begin{enumerate}
        \item It performs a HPO job for one or different modelling strategy/ies using the train\_model split. A strictly proper scoring rule is specified. See next Section \ref{sec:details_pl} for additional explanation.
        \item The best model according to the selected proper scoring rule is combined with $n$ calibrators to create $n$ potential models. The uncalibrated model is included as an option along with Platt, Isotonic and Spline calibration (See \ref{sec:app_des} for the details). The models are calibrated using the train\_calibration split.
    \end{enumerate}
    \item The ECCE metric of the potential models during the inference phase is estimated with K-fold Cross-validation.
    \item A model selector opts for the best potential model of the pool. In industry application analyzed,  ECCE-R is employed. See next Section \ref{sec:details_pl} for additional explanation about ECCE-R choice. See Section \ref{sec:app_des} to check how to selection is specifically done taking also the overall performance of the model  into consideration with a strictly proper scoring rule.
\end{enumerate}

\subsection{Additional details supporting the construction} \label{sec:details_pl}

HPO job of the model generator in Section \ref{sec:desc_pl} is done in function of a strictly proper scoring rule. The reason is that these rules offer comparative assessments of overall forecast quality \citep*{Brier1950,Gneiting2007}. In this way, we can select a base model that is good in both calibration (reliability) and discriminative power (resolution) and then try to improve the probability estimates with different calibrations. Examples of strictly proper scoring rules are the Brier Score and the Logarithmic score.  The Logarithmic score can be interesting when a high penalty wants to be imposed when the model is overconfident about an outcome and it fails. In fact, this metric can be very appealing, for instance,  when you want a high accuracy in a small range of low probabilities. Suppose the "true" probability is $0.01$ and you predict a probability of $0.0001$. The true probability is underestimated by a factor of $100$. This can be a big error in financial sectors such as debt collection or insurance and the log-loss can consider that. Imagine that you were predicting the probability of a car accident, the insurance company would be paying out $100$ times as much as they thought they would. On the other hand, the Brier score  is gentler than Logarithmic score and can be considered when the potential probability ranges are wider and the loss for the same error is preferably not dependent of this type of scales.

For the unfamiliar reader, a scoring rule is a function $S(\tilde{y},y)$ that assigns a loss or penalty in function of the probability estimate $\tilde{y} \in [0, 1]$ and the binary label $y \in \{0,1\}$. Infinite penalties can be allowed in the case where a prediction was declared to have probability zero and, consequently, be impossible. Furthermore, a scoring rule is proper if:
\begin{equation}
    \mathbb{E}[S(p,Y)] \leq \mathbb{E}[S(\tilde{y},Y)] \; \; \text{for all predicted} \; \; \tilde{y}
\end{equation}
where $Y$ is a Bernoulli random variable with success probability $p$. Moreover, it is strictly proper if the equality in the previous equation implies that $p = \tilde{y}$, so that the true success probability is the unique minimizer of the expected score.

Last but not least, the chosen ECCE criterion for the model selector of Section \ref{sec:desc_pl} is the ECCE-R. The reason is that it provides a bound of the total miscalibration over any interval of indices thus ensuring a global indicator of the calibrated estimates. In fact, the ECCE-R is the maximum of the absolute value of the total miscalibration over any interval of indices \citep*{ArrietaIbarra2022_metapaper}:
\begin{equation}
    \text{ECCE-R} = \max_{I} \mid  \sum_{j \in I} (\tilde{y}_j - y_j)/n \mid 
\end{equation}

\section{INDUSTRY APPLICATION AND RESULTS} 
\subsection{Description}\label{sec:app_des}

The purpose is create a model to give an useful propensity-to-pay score to the Operations Department of a debt collection company in order to optimize the strategy concerning amicable debt collection.  For instance, if the PtP score is high, a possible action is to reduce the call intensity to focus on more difficult cases. The list of actions taken based on the estimated probability is not indicated in this document to protect the company's strategy. 

In particular, the goal is to estimate the propensity to pay in the next seven days at the beginning of each week. The predictions will be done at debtor level. In particular, at the unique identifier of a group of debtors sharing debts.

The proposal in Section \ref{sec:ml_proposal} is implemented on a proprietary dataset, generously provided by an important debt collection company.
The data set contains 156 166 records. Each row of the dataset is uniquely identified by debtor identification numbers and a snapshot date. The data captures the time window from 2022-08 to 2023-04. Each row contains debtor and debt data at the snapshot date and a column indicating if the debtor paid after seven days. Features available are debt components (such as facevalue and interest amounts) and debtor behavior indicators (such as the number of payments done in the last months). The detailed exhaustive list cannot be given for the reasons in Section \ref{sec:data_available}. However, this is not an inconvenience to reproduce the proposed pipeline with the details of the previous Section and the specifics ones shown below.

Next, the configuration of the pipeline proposal is detailed.
\begin{enumerate}
\item Performance estimate with an outer 5-fold Cross-Validation
\item For each fold iteration, train\_model split represents 60 \% of data, train\_calibration represents 20 \% of data and test split represents 20 \% of data.
\item For this experiment, only LightGBM\footnote{See python package and documentation at \url{https://github.com/microsoft/LightGBM}} \citep*{ke2017lightgbm} is considered to accelerate the experiments (faster training speed, higher efficiency and lower memory usage). HPO job is carried out with the train\_model split with an inner cross-validation procedure (5-fold). The best candidate during the HPO job is selected according to the best cross-validated log-loss metric (strictly proper scoring rule). The training for a set of hyperparameters is performed using as objective function the binary log loss. If there is more than a modelling strategy to be considered (for instance, XGBoost \citep*{xgboost2016}), the model generator takes the best tuned models for both strategies and creates the pool of candidates with them.
\item Calibration is performed with the best model of the HPO job using the train\_calibration split. The calibration strategies considered are Platt Scaling, Isotonic Regression, Spline\footnote{Library Spline calibration \url{https://github.com/numeristical/introspective}. Other calibrations implemented using python scikit-learn package.} and uncalibrated. \citep*{Niculescu2005, Lucena2018}. Pipeline design is not limited to these procedures and other ones can be added if appropriate.
\item The ECCE metric for the Model selector is the ECCE-R for the reasons discussed in \ref{sec:details_pl}. In addition, a calibrated model will be chosen only if its log-loss is equal or lower than its uncalibrated version. This ensures that the calibrated version has at least the same overall performance than the uncalibrated one. This type of threshold and the chosen scoring rule can be adapted in function of the industry application. In the analyzed case, the log-loss scoring rule was selected to avoid overconfident wrong predictions that can have a big negative impact when, for instance, doing valuations based on the PtP forecast. And as previously introduced, the requested main focus is to improve calibration over other metrics because of the impact that it has in several business decisions based on the expected value estimates.
\end{enumerate}

\subsection{Results} \label{sec:results}

This section contain the results of the experiment described in Section \ref{sec:app_des}.

The model generator created four models to be evaluated: LightGBM, LightGBM + Platt Calibation, LightGBM + Isotonic Calibration and LightGBM + Spline Calibration. 

\begin{figure}[h] 
\centering
\includegraphics[scale=1.2]{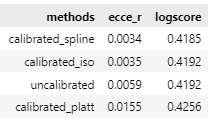}
\vspace{.3in}
\caption{ECCE-R metric and Logarithmic score of the models in the first performance estimation: LightGBM (uncalibrated), LightGBM + Platt Calibation (calibrated\_platt), LightGBM + Isotonic Calibration (calibrated\_iso) and LightGBM + Spline Calibration (calibrated\_ spline).}
\label{fig:summary_1ite}
\end{figure}

\begin{figure}[h] 
\centering
\includegraphics[scale=0.55]{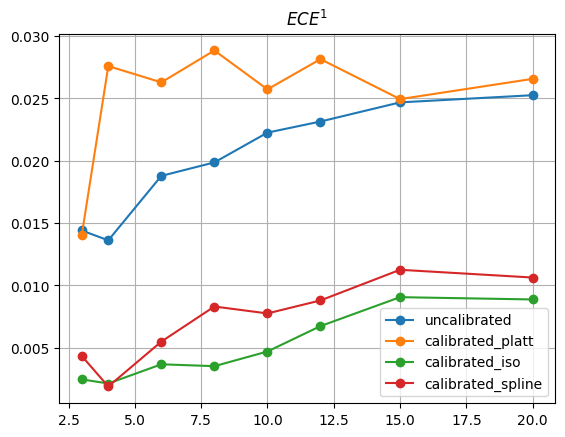}
\vspace{.3in}
\caption{$\text{ECE}^1$ in function of bins in the first performance estimation. Labels with the same meaning than Figure \ref{fig:summary_1ite}}
\label{fig:ece1_1ite}
\end{figure}

\begin{figure}[h] 
\centering
\includegraphics[scale=0.55]{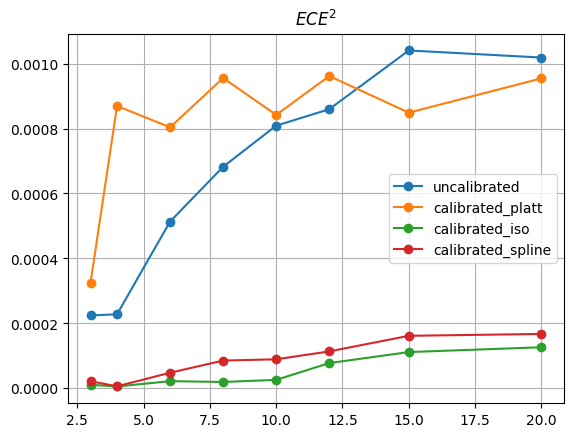}
\vspace{.3in}
\caption{$\text{ECE}^2$ in function of bins in the first performance estimation. Labels with the same meaning than Figure \ref{fig:summary_1ite}}
\label{fig:ece2_1ite}
\end{figure}

First, let's observe the partial ranking after the first iteration to observe the robustness of ECCE metrics over the widely used ECE for a model selector. This first iteration would be equivalent to do a performance evaluation with a single train-test split. Table \ref{fig:summary_1ite} shows the ranking at this stage with the performance estimates. We observe that the best model concerning both the Logarithmic Score and the ECCE-R is the LightGBM + Spline Calibration. If we compare the best score with the uncalibrated version, we can conclude that this Spline model has better overall performance and less miscalibration. In addition, the worst one would be clearly the Platt Calibration. However, if we compare the $\text{ECE}^1$ and $\text{ECE}^2$ metrics for an usual number of bins (20), the ranking is completely different, being the first one the Isotonic Calibration (e.g. 0.009  for $\text{ECE}^1$) and the last one the uncalibrated model (e.g. 0.025  for $\text{ECE}^1$). Figure \ref{fig:ece1_1ite} and \ref{fig:ece2_1ite} shows the evolution of the ECE metrics in function of the number of bins where we can see how the ranking can change erratically with them. According to the mathematical advantages detailed in Section \ref{sec:metrics} and proved in the mentioned references, the ECCE-R yield significantly more trustworthy results with less observations and with no parameters and thus it is preferred.

\begin{figure}[h] 
\centering
\includegraphics[scale=0.56]{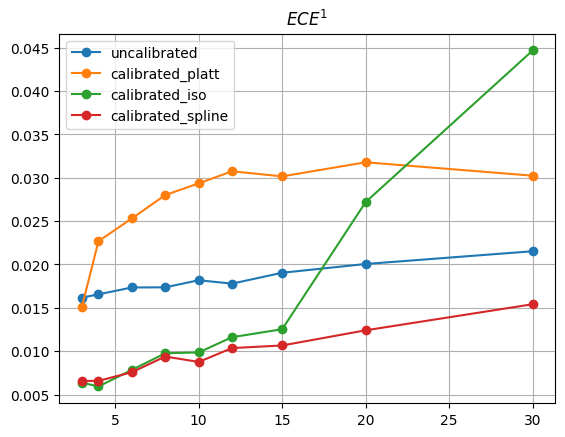}
\vspace{.3in}
\caption{$\text{ECE}^1$ in function of bins in the second iteration (second data split). Labels with the same meaning than Figure \ref{fig:summary_1ite}}
\label{fig:ece1_equiprob}
\end{figure}

Furthermore, these results are a practical example of how the unreliable ECE metrics can lead not only to select wrongly the best calibrated model but also to have potentially big mistakes or controversial decisions as saying that the Platt calibration is better than the uncalibrated model when we observe with more confidence the opposite. In fact, taking as ground truth the proved more reliable ECCE metric, the uncalibrated model has an error reduction in 61\% (with respect to Platt calibration) and not only that, but also a lower Logarithmic score. 

Moreover, the Figure \ref{fig:ece1_equiprob} is added as an additional example of calibration assessment in another scenario to illustrate better how fragile ECE metrics can be. Here, Isotonic calibration is estimated to be the best at the beginning, but when changing the number of bins from 15 to 20 the ECE evaluation changes dramatically, even being assessed as the worst when 30 bins are considered.

\begin{figure}[h] 
\centering
\vspace{.3in}
\includegraphics[scale=1]{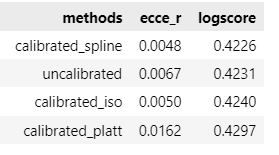}
\vspace{.3in}
\caption{Final ECCE-R metric and Logarithmic score of the models. Labels with the same meaning than Figure \ref{fig:summary_1ite}}
\label{fig:summary_cv}
\end{figure}

Next, the final performance estimates with 5-fold Cross-Validation for each potential model are detailed in table \ref{fig:summary_cv}.

According to the selection criterion discussed in Section \ref{sec:app_des}, the model selected is LightGBM + Spline Calibration: (the best ECCE-R with a log-loss equal or lower than its uncalibrated version). If we compare its score with the uncalibrated version, we can conclude that the final selected calibrated model have better overall performance and less miscalibration than the other options.

In addition, a experimentation phase was done in the company putting the model in production to be used in a limited subset of cases. The performances obtained where similar to the ones shown here, confirming in production that the model selected with the proposed pipeline was more reliable than the one selected with traditional ECE metrics.

It is interesting to mention that the strategy derived from operations with the new propensity-to-pay model has managed to increase in the experimental phase the collection success (estimated with a control and test group and bootstrapping the results). This last statement is not directly linked to the theoretical research but we think that it interesting to mention it in order to know that the models created with the proposal of the present work using the novel metrics proposed in the scientific research can yield positive results of this type to the industry.

\section{CONCLUSIONS} \label{sec:conclusions}
In this work, an advanced machine learning pipeline is proposed considering last advances and breakthroughs in the assessment of calibration to improve the propensity-to-pay models in the debt collection industry. We compared traditional metrics used in the industry (ECE metrics) to assess calibration with the recently proposed ECCE metrics where we showed practically the better reliability of the latest ones to perform model selection. Furthermore, the pipeline workflow is designed to address the application specifications, taking into account several criteria considering strictly proper scoring rules (in particular, the Logarithmic score) to avoid both model degradation of the overall performance and overconfident decisions that may damage the debt collection strategy.
Moreover, we showed that the proposal with the ECCE metrics is more convenient an robust for an automatic solution. This is a critical feature in the industry if you want to have a solution with minimal degradation (by constant updating), without manual intervention and without large human costs for constant evaluation.

\section{DATA AVAILABILITY STATEMENT} \label{sec:data_available}
Due to legal, commercial and confidentiality restrictions of the debt collection company, raw data cannot be shared. However, this is not important for presenting, discussing and reproducing the proposed methodology with the technical details showed and referenced in the paper. Moreover, a successful outcome is a sign that the proposed approach can address current industrial needs.

\bibliographystyle{unsrtnat}
\bibliography{refe_aistats2024}  

\end{document}